\documentclass[%
 aip,
 amsmath,amssymb,
preprint,%
]{revtex4-1}

\usepackage{graphicx}
\usepackage{dcolumn}
\usepackage{bm}

\usepackage[utf8]{inputenc}
\usepackage[T1]{fontenc}
\usepackage{mathptmx}
\usepackage{etoolbox}

\usepackage{xr}
\makeatletter
\newcommand*{\addFileDependency}[1]{
  \typeout{(#1)}
  \@addtofilelist{#1}
  \IfFileExists{#1}{}{\typeout{No file #1.}}
}
\makeatother

\newcommand*{\myexternaldocument}[1]{%
    \externaldocument{#1}%
    \addFileDependency{#1.tex}%
    \addFileDependency{#1.aux}%
}
\myexternaldocument{SPAD_Supplementary_Materials_APM}

\makeatletter
\def\@email#1#2{%
 \endgroup
 \patchcmd{\titleblock@produce}
  {\frontmatter@RRAPformat}
  {\frontmatter@RRAPformat{\produce@RRAP{*#1\href{mailto:#2}{#2}}}\frontmatter@RRAPformat}
  {}{}
}%
\makeatother
\begin{document}


\title{An efficient modeling workflow for high-performance nanowire single-photon avalanche detector}
\author{Zhe Li}
 \email{zhe.li@anu.edu.au;  lan.fu@anu.edu.au}
\author{H. Hoe Tan}
\author{Chennupati Jagadish}
\author{Lan Fu}
 \affiliation{Australian Research Council Centre of Excellence for Transformative Meta-Optical Systems, Department of Electronic Materials Engineering, Research School of Physics, The Australian National University, Canberra ACT 2601, Australia}

\begin{abstract}
Single-photon detector (SPD), an essential building block of the quantum communication system, plays a fundamental role in developing next-generation quantum technologies. In this work, we propose an efficient modeling workflow of nanowire SPDs utilizing avalanche breakdown at reverse-biased conditions. The proposed workflow is explored to maximize computational efficiency and balance time-consuming drift-diffusion simulation with fast script-based post-processing. Without excessive computational effort, we could predict a suite of key device performance metrics, including breakdown voltage, dark/light avalanche built-up time, photon detection efficiency, dark count rate, and the deterministic part of timing jitter due to device structures. Implementing the proposed workflow onto a single InP nanowire and comparing it to the extensively studied planar devices and superconducting nanowire SPDs, we showed the great potential of nanowire avalanche SPD to outperform their planar counterparts and obtain as superior performance as superconducting nanowires, i.e., achieve a high photon detection efficiency of 70\% with a dark count rate less than 20 Hz at non-cryogenic temperature. The proposed workflow is not limited to single-nanowire or nanowire-based device modeling and can be readily extended to more complicated two-/three dimensional structures.
\end{abstract}

\maketitle


\section{Introduction}

Single-photon detector (SPD), an essential building block for detecting and counting photons, plays a fundamental role in many emerging and fast-evolving technologies, such as quantum encryption and long-distance free-space quantum communication \cite{Hadfield_NatPhotonics_2009,Stajic_future_of_QM_Info_Processing,Pan_Nature_LightSciAppl_2015,Anisimova_etal_EPJ_QuanTech_2021}, light detection and ranging (LiDAR) \cite{Pan_etal_LiDAR_2021,Junsuk_etal_LiDAR_NatPhotonics_2021}, and single-photon time-of-flight 3D imaging \cite{Gupta_ICCV2019,Bruschini_SPD_imager_bio_2019,Piron_etal_review_SPAD_TOF_imaging_2021,Tobin_etal_real_time_3D_imaging_2021}, to name a few. Over the past two decades, significant efforts have been devoted to improving the performance of SPDs in terms of photon detection efficiency (PDE), dark count rate (DCR), timing jitter, and most importantly, the capacity of fully integrated generation, manipulation, and detection of single-photon quantum state all in one chip. Most commercially available SPDs operate based on the photomultiplier or avalanche mechanism to achieve single-photon level detection \cite{HuWeiDa_review_2022}.  Photon detection technologies based on the photomultiplier tube (PMT)  have long been developed and matured for almost a century. Although PMT offers a very large active area which is desirable for simplifying optical alignment in long-range free-space quantum communications, it requires kilovolt-level operating conditions and suffers from high dark count rate and low detection efficiency. SPDs based on avalanche mechanism have surpassed PMTs thanks to their lower operating voltage, higher photon detection efficiency, ease of integration into an array and compatibility with other parts of peripheral CMOS logic circuits. Indeed, Si and InGaAs/InP based single-photon avalanche detectors (SPADs) have dominated the market in the visible and near-infrared wavelength, respectively, with other semiconductor materials such as germanium-on-silicon, HgCdTe, InGaAs/InAlAs, and InSb sharing the rest \cite{Ceccarelli_SPAD_review_AdvQuanTech_2021,HuWeiDa_review_2022}. 

One of the main obstacles to advancing single-photon detection/counting technologies is that different performance metrics usually have conflicting trends (e.g. higher photon detection efficiency generally leads to larger dark count rate). Thus a trade-off in the design must be accounted for, depending on a specific application. This drives extensive research to explore new device structures made available by emerging low-dimensional nanostructures such as 1D nanowires (NWs) \cite{Esmaeil_SNSPD_2017,Vetter_SNSPD_2016,Korzh_SNSPD_2020,Wollman_SNSPD_2017} and 0D quantum dots \cite{Eisaman_review_2011,Blakesley_QD_RT_SPD,Gansen_QD_FET_SPD_2007}. For example, it has been demonstrated that superconducting nanowire single-photon detectors (SNSPDs) promise the potential to closely approach an ideal SPD performance by simultaneously achieving low dark count rate (1 Hz), high detection efficiency (90\% at wavelengths between 1520-1610 nm) and low timing jitter (150 ps) \cite{Marsili_SNSPD_NatPhotonics_2013}. However, a critical bottleneck of SNSPD devices is its requirement of cryogenic temperature ($<$4 K), which drastically adds to the cost and complexity of the experimental setup and becomes prohibitive to use for applications such as satellite optical receivers \cite{Ceccarelli_SPAD_review_AdvQuanTech_2021}.

On the other hand, exploring 1D semiconductor nanowire-based single-photon avalanche detectors (NW-SPADs) has drawn increasing attention recently. It promises to deliver as good performance as SNSPDs, i.e. high photon detection efficiency and low timing jitter, but operating near room temperature \cite{Farrell_Huffaker_InGaAs_GaAs_SPAD_nanoLett_2019}. Furthermore, the unique advantage of NW-SPAD is that low dark count rate performance can be maintained despite operating near room temperature, thanks to its nanoscale size and high material quality \cite{QianGao_InP_NW_review_2019}. Over the past decade, although many linear-mode NW-APDs have been reported \cite{Hayden2006,Yang_Lieber_Si_NW_APD_NanoLetts_2006,Yang_etal_Si_NW_APD_2014,Chuang_etal_GaAs_NW_APD_NanoLetts_2011,Bulgarini_SingleNW_QD_APD_NatPhotonics_2012,Jain_etal_InP_SAM_APD_2017,Senanayake_etal_GaAs_NW_APD_NanoLetts_2012,Farrell_GaAs_NW_APD_2015,Parakh_etal_GaAsSb_GaAs_NW_APD_2022,Pokharel_etal_GaAs_GaAsSb_NW_APD_2023}, there is only one report on nanowire APDs operating above their breakdown voltage (i.e., NW-SPADs or Geiger-mode NW-APDs) \cite{Farrell_Huffaker_InGaAs_GaAs_SPAD_nanoLett_2019}, because of the stringent requirement in epitaxial growth and device fabrication of these types of nanowire devices. It requires prudent device structural design and numerous cycles of complicated and (perhaps) non-reproducible device fabrication, which is costly and time-consuming. Hence, there is a great demand for developing a comprehensive device modeling scheme to achieve a deep understanding of the critical design parameters and provide timely guidance to material growth and device fabrication. 

Over the past two decades, many numerical studies have been dedicated to modeling SPAD performance \cite{Spinelli_SPAD_Physics_and_Modeling_IEEE1997,Donnelly_SPAD_design_consideration,XudongJiang_etal_IEEE_2007,Xu_SPAD_SilvacoModeling_2016,Ahammed_etal_SPAD_modeling2017,Yanikgonul_SPAD_modeling_IEEE2020}. The seminal work by Spinelli and Lacaita \cite{Spinelli_SPAD_Physics_and_Modeling_IEEE1997} pursued both deterministic (i.e. drift-diffusion model \cite{Sze_PhysofSemiDev_3rd_ch2}) and stochastic (simplified Monte-Carlo method \cite{Plimmer_SimpleMonteCarlo_IEEE1999}) modeling techniques to simulate transient terminal avalanche current behavior and timing resolution. However, as pointed out by the authors, fully time-dependent deterministic modeling required a formidable computational load with serious numerical problems \cite{Laux_SPAD_numerical_issue_1985}. Thus the majority of the work relied on simplified and equivalent lumped models. Other reported work either focused on one or two aspects of SPAD performance or adopted a simplified model to mitigate the heavy computational load of simulating avalanche phenomena. For example, \citeauthor{Donnelly_SPAD_design_consideration} and \citeauthor{XudongJiang_etal_IEEE_2007} studied extensively the trade-offs between dark count rate and photon detection efficiency\cite{Donnelly_SPAD_design_consideration,XudongJiang_etal_IEEE_2007}; however, a complete drift-diffusion modeling was not considered. Overall, there is still a lack of robust and efficient computational workflow to simulate a suite of key SPAD performance metrics in the field. 

In this work, we propose a hybrid numerical workflow that integrates time-dependent drift-diffusion (DD) modeling, steady-state DD modeling, as well as a range of auxiliary post-processing routines to predict the five most important SPAD performance metrics, i.e., avalanche breakdown voltage, avalanche built-up time (corresponding to SPAD sensing time), timing jitter, dark count rate, and photon detection efficiency. Although the proposed workflow was demonstrated with single-nanowire SPAD in a one-dimensional domain, it is general to more complicated structures in two/three dimensions and not limited to nanowire-based devices. It can also be extended to include more SPAD metrics, such as afterpulsing probability, in future work.

\section{Numerical methods}
\subsection{Proposed workflow}
\begin{figure*}[h]
  \includegraphics[width=1\textwidth]{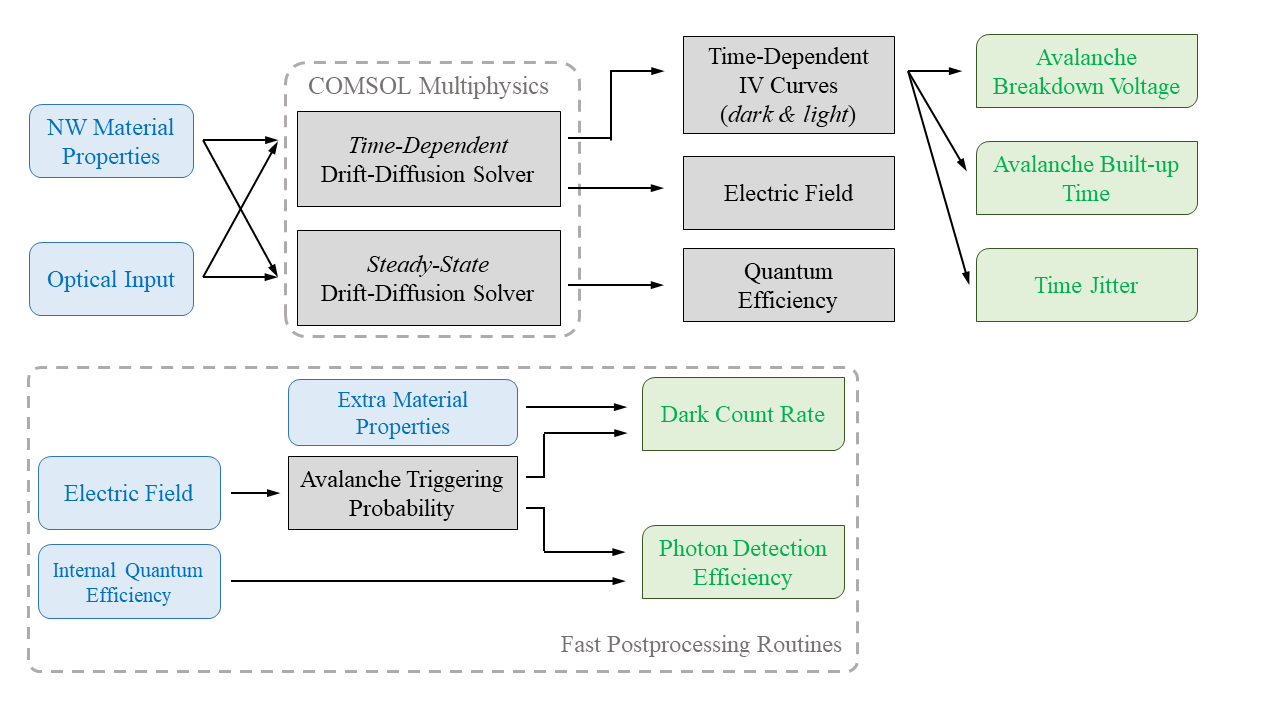}
  \caption{Proposed workflow for simulating a suite of SPAD performance metrics (in green), including avalanche breakdown voltage, avalanche built-up time (from dark time-dependent IV curves) and timing jitter (from light time-dependent IV curves), which are directly generated from the DD solver. Post-processed metrics include dark count rate and photon detection efficiency, computed based on the electric field and quantum efficiency generated by DD solvers. The input parameters are depicted in blue. }
  \label{fig:workflow_diagram}
\end{figure*}

To maximize the usage of time-consuming drift-diffusion simulation and produce as many device performance metrics as possible, we proposed the workflow shown in Fig.~\ref{fig:workflow_diagram}, to predict the five most important SPAD metrics: avalanche built-up voltage, avalanche built-up time, timing jitter, dark count rate (DCR) and photon detection efficiency (PDE). The COMSOL Multiphysics was employed for the drift-diffusion (DD) simulations in this work. The workflow consists of two parts: a relatively slow but comprehensive time-dependent DD solver that produces three out of five SPAD metrics and other intermediate results for the subsequent process; and a fast post-processing routine that takes some of the output from DD simulation as input to compute the rest of performance metrics. 

More specifically, the time-dependent DD solver is set up to compute time-dependent current-voltage (IV) characteristics under dark and light conditions with various external biases. Based on the IV curves, information on breakdown voltage can be immediately extracted. For applied bias beyond avalanche breakdown, terminal current increases rapidly and reaches a plateau after an extended period (e.g., see Fig.~\ref{fig:time_dependent_IV_and_terminal_current}(b)). By fitting a standard logistic function to the current-time curve in dark condition, the avalanche built-up time can be estimated, which corresponds to the lower limit of SPADs' sensing time (its inverse gives the maximum count rate). Likewise, the light avalanche built-up time can be calculated under an optical excitation. However, such response time varies depending on which part of the NW is under light excitation, and the corresponding standard deviation estimates the (lower limit of) timing jitter. 

Deploying a time-dependent DD solver provides important metrics of interest and ensures better numerical stability and convergence than the commonly used steady-state DD solver. This is because the regime after avalanche breakdown involves high non-linearity and exponential increase of carriers in a very short period, which is by no means a steady-state condition. Moreover, the space-charge effect that limits avalanche current from approaching infinity cannot be correctly simulated with a steady-state DD scheme, which is precisely the reason that steady-state DD simulations always tend to diverge just a few volts beyond breakdown \cite{failure_example_Lumerical_APD}. The examples of the electric field with space-charge effect correctly recovered after avalanche breakdown can be found in Fig.~\ref{fig:efield} in the supplementary material. Therefore, one must apply a time-dependent DD solver (within the framework of deterministic modeling of SPADs) to accurately predict SPAD device behavior, even though the apparent downside is that time-dependent DD simulations are very time-consuming, which accounts for over 90\% of the total computation time.  

On the other hand, although the steady-state DD scheme cannot accurately simulate avalanche phenomena, it can still compute quantum efficiency (QE), which characterizes how good incident photons are absorbed in NWs, regardless of whether there is an avalanche. This is a much faster routine and has been used for NW solar cell simulations with negligible computing time because it only requires a short-circuit condition (i.e. zero bias) \cite{LI_NWSC_IEEEJPV_2015,ZHONG2016106,Inseok_NW_LED_NanoLett_2019}. The generated QE is an input for computing photon detection efficiency, as discussed in the following section. 

A fast post-processing routine is set up outside the drift-diffusion simulation workflow, taking the intermediate results from the drift-diffusion simulations as input and directly computing the dark count rate and photon detection efficiency. The spatial distribution of the electric field given by the time-dependent DD simulation is used to calculate the avalanche triggering probability (ATP), the chance that an electron-hole pair can trigger an avalanche event, following the scheme proposed by \citeauthor{McIntyre_1973_ATP} \cite{McIntyre_1973_ATP}. For SPADs, it is more meaningful to cite photon detection efficiency \cite{Mazzillo_PDE_cal_IEEE2008,Gulinatti_Cova_SPAD_DD_MonteCarlo_2009,Panglosse_modeling_PDE_CMOS_SPAD_2021,Qian_review_modeling_SPAD_2023}, defined as the product of quantum efficiency and avalanche triggering probability. This is calculated by the post-processing routine. 

The dark count rate is another critical performance metric that characterizes the SPAD miscounts in the absence of a target optical signal. It indicates the occurrence of avalanche events triggered by thermally generated carriers or through other mechanisms such as direct or trap-assisted tunneling in the multiplication region, also known as dark carrier generation. These carrier generation terms are strong functions of the local electric field and depend on material properties, especially the material defect density and energy levels. The fast routine can easily compute these carrier generation rates by combining the electric field (extracted from DD simulations) and information on defects. The dark count rate is simply a product of dark carrier generation and avalanche triggering probability, given that every dark-generated carrier only has a certain chance to trigger an avalanche event. 


Overall, the proposed workflow seeks to maximize the usage of time-consuming DD simulation; both direct and indirect solutions from time-dependent DD simulations are used to calculate key SPAD performance metrics. 

\subsection{Modeling SPAD performance}\label{sec:modeling_SPD_performance}

The time-dependent drift-diffusion model, which solves for IV curves and obtains spatial electric field, consists of three coupled differential equations: the continuity equations for electrons and holes, and the Poisson equation for electric potential \cite{Sze_PhysofSemiDev_3rd_ch2}. To demonstrate the proposed workflow and highlight the potential of nanowire-based SPADs, all simulations in this work were performed at a temperature of 150 K, higher than the cryogenic threshold temperature (120 K) \cite{lagowski1997macmillan}. The details of temperature dependent material properties can be found in the supplementary material. Moreover, since the working regime of Geiger-mode SPADs involves a large electric field after avalanche breakdown, the high-field saturation of carrier drift velocity must be considered. We adopt the commonly used Caughey-Thomas field-dependent model for carrier mobility \cite{Caughey_Thomas_mobility_model_1967}, also detailed in the supplementary material. 

To accelerate the time-dependent DD simulation, we include only the most essential physics relevant to avalanche SPAD devices: for carrier recombination, the standard Shockley-Read-Hall model is applied\cite{Sze_PhysofSemiDev_3rd_ch2} ; for carrier generation, impact ionization and optical generation are considered. In particular, we adopt the ionization coefficient model proposed in Ref.~\citenum{Petticrew_2020_modeling_II_InP} based on Monte Carlo simulations, verified by experimental data for temperatures ranging from 150 to 290 K. For the light generation model, for simplicity, we assumed a laser source with single-photon level power that gives a spatially uniform carrier generation rate inside the NW, as the light signal used in this work is mainly for computing timing jitter. Optical simulation (e.g., finite-difference time-domain (FDTD) simulation) can also be performed to obtain a more realistic carrier generation profile, though not critical for this study \cite{LI_NWSC_IEEEJPV_2015,ZY_Li_Zahra_InAs_InfraredPD_2023}.

The terminal currents versus transient time under different external biases are a direct product of time-dependent DD simulation (e.g., Fig.~\ref{fig:time_dependent_IV_and_terminal_current}(b)), from which breakdown voltage can be immediately determined. The dark avalanche built-up time is estimated by fitting a logistic function to the terminal current. Similarly, when optical generation is added, light current versus time can be obtained. The avalanche built-up time under light varies along the NW as it depends on which part of NW is excited (assuming a laser excitation with finite spot size). We can simulate a series of time-current curves by scanning a finite spot-size light source along the NW, and thus obtaining a range of avalanche built-up time. The built-up time here includes the time it takes to reach a threshold impact-ionization electric field, the time for carriers generated in a low-field region drifting to the high-field multiplication region, and the time for carriers generated in quasi-neutral region diffusing towards the multiplication region. In reality, all three above processes involve statistical fluctuation, incurring an extra and random delay. However, this work aims at fast deterministic modeling of SPADs, so these stochastic processes are not included. Therefore, the resultant built-up time is a lower limit of a SPAD's timing response, i.e., theoretically, the best sensing time. Furthermore, the standard deviation of light avalanche built-up time estimates the SPAD timing jitter (also being a lower limit). Lateral diffusion of carriers in the direction perpendicular to the external bias is not considered, which needs an extension to the 2D/3D simulation domain. Such lateral diffusion within the depletion region is expected to be minimal since the high electric field leads to drift-dominant transport. The diffusion in the quasi-neutral region depends on the nanowire radius and will be investigated in future work, especially when surface recombination is also introduced. 

Dark count rate (DCR) and photon detection efficiency (PDE) are calculated via post-processing routines, both of which rely on the electric field extracted from time-dependent DD simulations. The electric field as a function of external biases is used to compute impact ionization coefficients $\alpha_e$ and $\alpha_h$ for electrons and holes, respectively, defined in Ref.~\citenum{Petticrew_2020_modeling_II_InP}, which strongly depend on the spatial distribution of the local electric field. These coefficients play a crucial role in determining the avalanche triggering probability (ATP), which characterizes the probability of generated electron-hole pair successfully triggering a self-sustaining avalanche, and is thus needed for calculating DCR and PDE. We denote this probability as $P_{pair}$:
\begin{equation}
    P_{pair}(x) = P_e(x) + P_h(x) - P_e(x) P_h(x),
\end{equation}
where $P_e$ and $P_h$ are the probabilities that an electron or a hole can trigger an avalanche event, respectively; the subtraction of $P_e P_h$ in the equation is to avoid counting twice the avalanche triggered by both an electron and a hole \cite{Ceccarelli_SPAD_review_AdvQuanTech_2021}. It has long been proposed that $P_e$ and $P_h$ can be modeled through the following differential equations for a p-i-n structure, where the p-type region starts from $x=0$ and ends with the n-type region at $x=L$ with $L$ being the length of an NW \cite{Oldham_etal_1972_ATP_Original}:

\begin{subequations}
    \begin{align}
    \frac{\partial P_e}{\partial x} &= -(1-P_e) \alpha_e (P_e+P_h-P_e P_h), \\
    \frac{\partial P_h}{\partial x} &= (1-P_h) \alpha_h (P_e+P_h-P_e P_h).
\end{align}
\end{subequations}

Following the scheme proposed in Ref.~\citenum{McIntyre_1973_ATP}, solutions of coupled equations above can be found by solving a transcendental equation for $P_e(x=0)$:
\begin{equation}
	1 = \left[1-P_e(0) \right]\exp \left[\int_0^L \frac{\alpha_e(x')P_e(0)f(x')}{1-P_e(0)+P_e(0)f(x')}dx' \right],
\end{equation} 
given the boundary condition that $P_e(L)=P_h(0)=0$, and $f(x)$ is defined as:
\begin{equation}
	f(x) = \exp \left\{\int_0^x \left[\alpha_h(x')-\alpha_e(x') \right] dx' \right\}.
\end{equation}
Once $P_e(0)$ is solved, the ATP for electron-hole pairs, electrons and holes, respectively, are given by:
\begin{subequations}
\begin{align}
    P_{pair}(x) &= \frac{P_e(0)f(x)}{1-P_e(0)+P_e(0)f(x)}, \\
	P_e(x) &= 1-\left[1-P_e(0)\right]\exp \left[\int_0^x \alpha_e(x') P_{pair}(x') dx' \right],\\
	P_h(x) &= \frac{P_{pair}(x)-P_e(x)}{1-P_e(x)}.
\end{align}
\end{subequations}
With the solved avalanche triggering probability $P_{pair}$, the DCR can be determined by:
\begin{equation}
	\mbox{DCR} = A \int_0^L P_{pair}(x) G_{\scriptsize{\mbox{tot}}}(x) dx,
\end{equation}
where $A$ is the cross-sectional area of the NW, and $G_{\scriptsize{\mbox{tot}}}$ is the total dark carrier generation rate. In this work, we include contributions from thermal generation based on the Shockley-Read-Hall model ($G_{\scriptsize{\mbox{SRH}}}$), direct band-to-band tunneling ($G_{\scriptsize{\mbox{BBT}}}$), and trap-assisted tunneling ($G_{\scriptsize{\mbox{TAT}}}$) \cite{Donnelly_SPAD_design_consideration,XudongJiang_etal_IEEE_2007,Xu_SPAD_SilvacoModeling_2016}; the corresponding formulations can be found in the supplementary material. 

As mentioned before, extra information on quantum efficiency (QE) that characterizes photon absorption and carrier collection is needed to compute photon detection efficiency. The spatial QE profile can be calculated as: 
\begin{equation}\label{eq:Pabs}
	\mbox{QE}(x) = \frac{I_{\scriptsize{\mbox{ph}}}(x)/q}{G_{\scriptsize{\mbox{opt}}} 
 \mathcal{V}_{\scriptsize{\mbox{opt}}}},
\end{equation}
where $I_{\scriptsize{\mbox{ph}}}$ is the photo-current measured from the device terminal, whose magnitude depends on which part of the NW is optically excited (thus a function of $x$). $G_{\scriptsize{\mbox{opt}}}$ is the carrier generation rate due to the optical signal, and $\mathcal{V}_{\scriptsize{\mbox{opt}}}$ is the volume where $G_{\scriptsize{\mbox{opt}}}$ is applied, which can be approximated as $\mathcal{V}_{\scriptsize{\mbox{opt}}}\approx A d$, where $d$ is the optical excitation size (i.e. averaged beam size); here $d$ can be set to a realistic value based on the laser spot size, or it can be smaller if a detailed mapping of QE is desired to fully characterize its optical response. The light carrier generation $G_{\scriptsize{\mbox{opt}}}$ is related to material absorption, reflection, incident laser power, and device geometries. For array-based nanowire devices with sub-wavelength feature sizes, the resonant light trapping renders the conventional ray optics description inaccurate. In that case, $G_{\scriptsize{\mbox{opt}}}$ has no simple analytical formulation, such as the Beer-Lambertian type of absorption commonly adopted in the previous works \cite{Mazzillo_PDE_cal_IEEE2008,Gulinatti_Cova_SPAD_DD_MonteCarlo_2009,Panglosse_modeling_PDE_CMOS_SPAD_2021,Qian_review_modeling_SPAD_2023}. As a result, a more complicated light absorption profile $G_{\scriptsize{\mbox{opt}}}$ can only be obtained by numerical simulations \cite{Hu_Chen_LightTrapping_SiNWSC_NanoLetts_2007,Kupec_LightTraping_SiNWSC_OE2010,Anttu_Xu_LightTrapping_NWSC_OE2013}. Nevertheless, performing extra numerical simulations to obtain a more realistic $G_{\scriptsize{\mbox{opt}}}$ for a single-nanowire device is not essential for this study. Thus, we assumed a spatially uniform generation rate and adopted the following simple formulation:
\begin{equation}
	G_{\scriptsize{\mbox{opt}}} (\lambda) = \alpha(\lambda) \left[1-R(\lambda) \right] \frac{P_{\scriptsize{\mbox{Laser}}}}{d^2} \frac{\lambda}{h c}, 
\end{equation}
where $P_{\scriptsize{\mbox{Laser}}}$ is the laser power, $\alpha(\lambda)$ is the absorption coefficient, $R(\lambda)$  is the fraction of incident photons reflected from the NW surface (considered negligible in this work), $c$ is the speed of light in vacuum, $h$ is the Planck's constant, and the area of incident laser on NW surface is approximated as $d^2$.

The spatially resolved photon detection efficiency is then given by:
\begin{equation}\label{eq:spatial_PDE}
    \mbox{PDE}(x) = \mbox{QE}(x) \times P_{pair}(x),
\end{equation}
where it is noted that in the low-field or quasi-neutral region, the avalanche triggering probability $P_{pair}$, due to its definition, reduces to either $P_e$ or $P_h$. A single-valued PDE is sometimes more convenient to cite for performance comparison among different device designs. We thus defined a full-width at half-maximum length-averaged PDE$(x)$ as:
\begin{equation}\label{eq:length_avg_PDE}
    \langle \mbox{PDE} \rangle = \frac{1}{x_2-x_1} \int_{x_1}^{x_2} \mbox{PDE}(x) dx,
\end{equation}
where $x_1$ and $x_2$ define the boundary $x$ coordinates that cover the half maximum. 

\section{Results and discussion}
\begin{figure*}[h]
\includegraphics[width=1\textwidth]{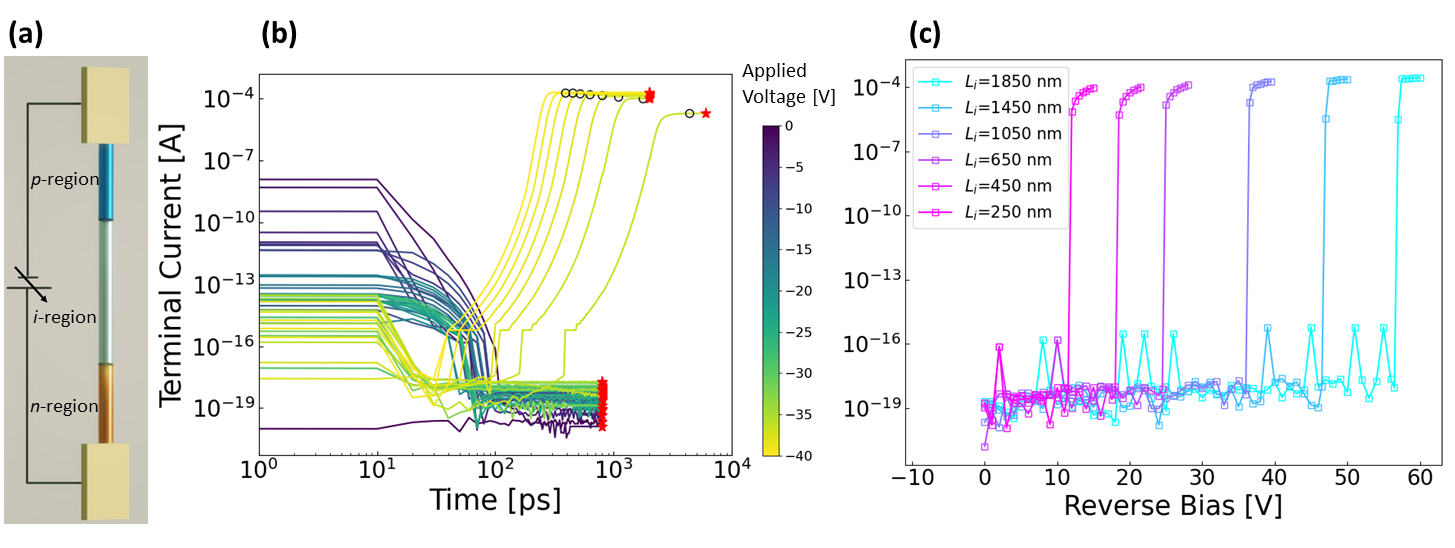}
\caption{(a) A simplified schematic of single-NW based SPAD. (b) An example of time-dependent current characteristics in the dark condition for i-region thickness of 950 nm with p-/n- and i-segment doping of $1\times 10^{18}$ and $5\times 10^{15}$ cm$^{-3}$, respectively. The black circle indicates avalanche built-up time calculated by fitting a logistic function to the time-current curve; the red star highlights the end of the simulation where the final terminal current is extracted. (c) Current-voltage characteristics extracted from time-dependent current curves for a range of i-region thickness ($L_i$) with the same doping as in (b).
}  
\label{fig:time_dependent_IV_and_terminal_current}
\end{figure*}

To investigate the performance of single-NW based SPAD, a p-i-n homojunction indium phosphide (InP) NW lying horizontally on a substrate is simulated. An Ohmic contact is assumed at the ends of both p- and n-segments. The NW has a length of 3 $\mu$m, with the p-segment length fixed at 1 $\mu$m and the i-segment varied from 250 to 1850 nm, whilst the n-segment length is varied accordingly to give a fixed total NW length. A simplified device schematic is shown in Fig.~\ref{fig:time_dependent_IV_and_terminal_current}(a). Assuming that NW is grown from an n-type InP substrate, a highly doped n-type InP segment can be grown first, followed by an unintentionally doped i-region and p-doped top segment. In practice, it is still challenging to have a heavily doped p-type NW by selective-area metal-organic vapor-phase epitaxy (SA-MOVPE)  \cite{QianGao_InP_NW_review_2019}. Hence, based on the experimental information, we fixed the n-segment doping at $1\times10^{18}$ cm$^{-3}$ and varied the p-type doping from $1\times10^{17}$ cm$^{-3}$ (experimentally achievable  \cite{Ziyuan_IIIV_SingleNW_SolarCell_Review_2018,QianGao_InP_NW_review_2019}) to $1\times10^{18}$ cm$^{-3}$. The i-region is slightly n-doped due to the background doping often observed in undoped InP NWs by SA-MOVPE growth \cite{QianGao_InP_NW_review_2019}. The i-region doping is also varied from $5\times10^{15}$ to $1\times10^{17}$ cm$^{-3}$ to explore its impact on SPAD performance. 

Solving the time-dependent DD model gives the time-dependent terminal current at different biases as direct output. An example is shown in Fig.~\ref{fig:time_dependent_IV_and_terminal_current}(b) for an i-region thickness of 950 nm with p-/n- and i-region doping of $1\times10^{18}$ and $5\times10^{15}$ cm$^{-3}$, respectively. The temporal behavior of terminal current below and beyond avalanche breakdown can be easily seen. Below the breakdown, the current experience some initial fluctuation and gradually reaches a steady state with a negligibly small current. On the other hand, as applied bias becomes equal to or greater than the breakdown voltage, the current rapidly ramps up and soon saturates at a certain level. The black circle highlights the avalanche built-up time, which is the period elapsed from the external bias is applied until a stable avalanche current is obtained. To estimate such built-up time, the time-current curve is first normalized by its maximum current and fitted by a logistic function. Truncating the fitting curve where its gradient is smaller than a prescribed threshold determines the build-up time. We take 10$^{-6}$ ps$^{-1}$ as the truncation threshold, considered small enough to be a reasonable criterion for a steady state. The fact that time-dependent current reaches a plateau when avalanche breakdown occurs indicates that the space-charge effect is correctly recovered in the simulations. Hence the current does not tend to go to infinity (see more discussion in the supplementary material).

Once the breakdown current reaches a steady state, the terminal current at the last simulation timestep is extracted as the final terminal current, usually a few hundred picoseconds longer than the avalanche built-up time, highlighted in red stars in Fig.~\ref{fig:time_dependent_IV_and_terminal_current}(b). No further simulation beyond this point is needed since the current fluctuation is too small to detect experimentally. Fig.~\ref{fig:time_dependent_IV_and_terminal_current}(c) shows the current-voltage characteristics for a range of i-region thicknesses $L_i$ with the same doping as in Fig.~\ref{fig:time_dependent_IV_and_terminal_current}(b). It can be clearly seen that the breakdown voltage reduces with the decrease of i-region thickness.

\begin{figure}[h]
\includegraphics[width=0.8\textwidth]{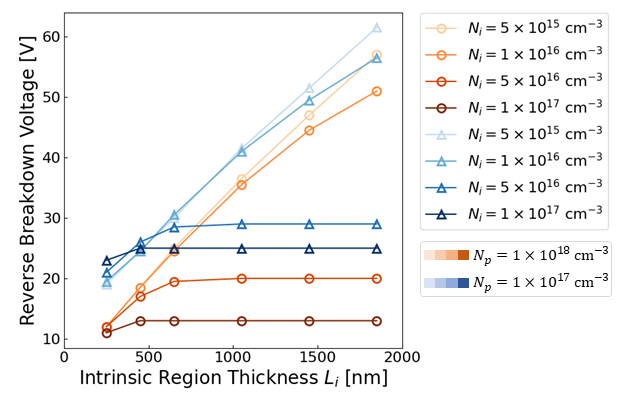}
\caption{Avalanche breakdown voltage vs i-region thickness for various p-/i-segment doping cases.
}  
\label{fig:Vb_vs_i_region_thickness}
\end{figure}

Fig.~\ref{fig:Vb_vs_i_region_thickness} summarizes the extracted breakdown voltage as a function of i-region thickness for all doping cases, showing that a thinner i-region gives a smaller breakdown voltage. Avalanche breakdown is triggered once the electric field within the i-region reaches a threshold. For a thinner i-region, a smaller bias is enough to reach the threshold electric field. Although the breakdown voltage for a particular i-region thickness is similar between nanowires and the planar counterparts (e.g., see Ref.~\citenum{LJJ_Tan_Planar_InP_APD}), it does not necessarily cause other issues, such as comparably high dark count rates, thanks to the nanowire's miniaturized volume, which will be discussed later. For the same p-segment doping (grouped by red or blue grading colors) and i-region thickness, an increase in i-region doping also gives a smaller breakdown voltage, consistent with widely cited theoretical analysis \cite{Sze_PhysofSemiDev_3rd_ch2}. Overall, higher p-segment doping reduces breakdown voltage for all i-region thickness/doping cases. Moreover, as i-region doping increases and becomes comparable to p-region doping (e.g., $N_i\geq 5\times10^{16}$ cm$^{-3}$), the breakdown voltage is less dependent on i-region thickness since the effective multiplication region is primarily located at the p-i junction and does not span the entire i-region. Hence, varying i-region thickness has a limited impact on the breakdown voltage.

\begin{figure*}[h]
\includegraphics[width=1\textwidth]{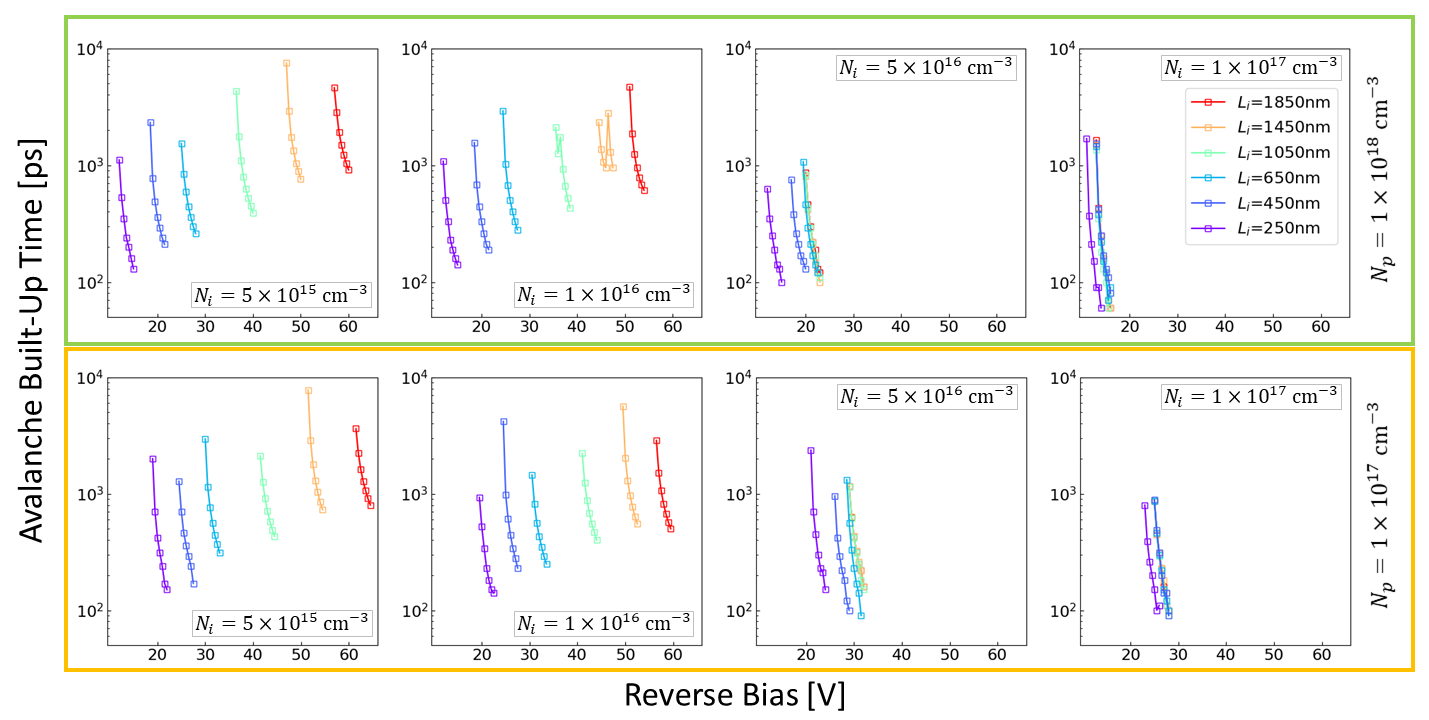}
\caption{Avalanche built-up time extracted from dark time-dependent current curves for different i-region thicknesses and p-/i-segment doping. The top row corresponds to a varying i-region doping with the p-segment doping of $1\times10^{18}$ cm$^{-3}$, and the bottom row is for the p-segment doping of $1\times 10^{17}$ cm$^{-3}$. 
}  
\label{fig:ava_builtup_time}
\end{figure*}

The avalanche built-up time extracted from the dark time-dependent current curve is plotted against reverse bias beyond breakdown, as shown in Fig.~\ref{fig:ava_builtup_time} for all doping cases. This `dark' avalanche built-up time can be regarded as the lower bound of the SPAD sensing response, as it characterizes the time taken to trigger an avalanche breakdown without photon excitation. Under the light condition, the response time will usually be shorter (with a few exceptions, as discussed later). However, there is no longer a single-valued built-up time to cite in the light case, as it depends on which part of the NW is optically excited. The inverse of avalanche built-up time can also be seen as the intrinsic maximum count rate of photons, being `intrinsic' because it is independent of the effects from quenching, hold-off and reset times that usually depend on peripheral quenching circuits and specific applications (but not device itself). 

We performed simulations up to 3 V above avalanche breakdown voltage for all testing cases. Fig.~\ref{fig:ava_builtup_time} shows that avalanche built-up time drops quickly by one order of magnitude when reverse bias increases by only several volts. Regardless of the i-region thickness, for both high and low p-segment doping cases, the dark built-up time can drop below 1 ns when the reverse bias is 1.0 V beyond breakdown. For p- and i-region doping of $10^{18}$ and $10^{16}$ cm$^{-3}$, respectively, the built-up time fluctuates for $L_i = 1050$ and 1450 nm. This is because some of the current experience a sudden surge, quickly reaching a comparable current level to other $L_i$ cases. However, the current is yet to stabilize, resulting in a longer total built-up time. Overall, a larger reverse bias is preferred to achieve a sub-nanosecond SPAD sensing time and a better maximum count rate. However, a larger bias with high electric field in the i-region also leads to a higher dark count rate. Such a trade-off must be considered, depending on whether the maximum count rate or DCR poses a more stringent limit on specific applications.

\begin{figure*}
\includegraphics[width=1\textwidth]{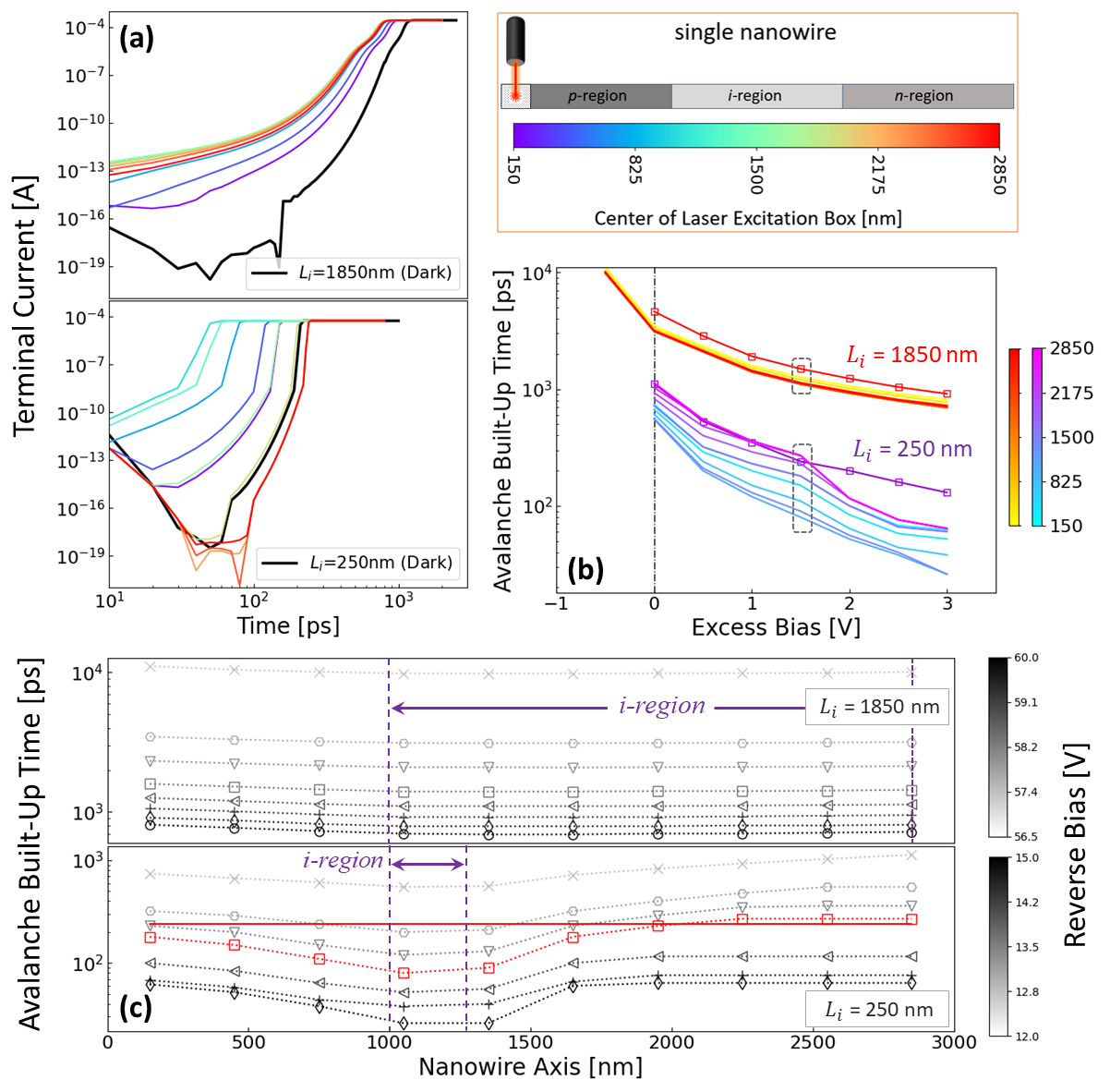}
\caption{ Avalanche breakdown of NW-SPAD under light condition. (a) At $V_e=$1.5 V, time-dependent terminal currents are extracted for the cases of $L_i=$ 1850 nm (upper panel) and 250 nm (lower panel), with p-/i-doping of $10^{18}$ and $5\times 10^{15}$ cm$^{-3}$, respectively. The colored data correspond to light response with a laser beam size of $300\times300$ nm$^2$) scanned from the p- to n-side of the NW, as indicated in the upper right schematic. (b) Both dark (line plus symbol) and light (line only) avalanche built-up time are summarized at different excess biases, with different colormaps representing laser scanning locations. Data highlighted in the dashed frame correspond to the cases shown in (a). (c) Spatial variation of light avalanche built-up time is plotted along the NW axis at different reverse biases. For $L_i=$250 nm at a reverse bias of 13.5 V(i.e. $V_e$=1.5 V), the data is plotted in red with its dark counterpart also shown in a red line. 
}  
\label{fig:light_ava_builtup_time}
\end{figure*}

When the device is exposed to optical excitation, the light time-dependent terminal currents can similarly be simulated to extract the light avalanche built-up time. The single-photon source used in this work has an input power of 5.68 pW at a wavelength of 700 nm, which gives a single photon per pulse, assuming a laser repetition rate of 20 MHz commonly used in experiments \cite{ZhuYi_SelfPowered_SPD_AM_2021}. The top right schematic in Fig.~\ref{fig:light_ava_builtup_time} shows the setup of simulations in the light condition: a laser spot size of 300 nm in diameter is assumed, within which the carrier generation is enabled. Here the laser spot size is much smaller than a realistic laser spot size because we want to obtain a detailed mapping that characterizes the NW intrinsic photo-response, which in fact, should be independent of optical excitation spot size. Initially, the excitation `box' is placed at the p-doped end of NW, and is progressively scanned through the entire NW with a step size of 300 nm. In total, it generates ten light simulation data, each giving a light avalanche built-up time. The light avalanche built-up time varies due to an interplay of several competing processes (i.e., light-generated carrier diffusion, low-field drift and recombination). The variation of spatially dependent light avalanche built-up time thus induces a timing jitter. It should be noted that such timing jitter is a lower limit of the actual one since various statistical fluctuations of avalanche are not considered; in other words, it is the `deterministic' part of the actual timing jitter, which can be well predicted in this proposed workflow once the device structure is given. 

Fig.~\ref{fig:light_ava_builtup_time}(a) shows two examples of time-dependent terminal currents under light condition for i-region thickness of 1850 and 250 nm, respectively, both with p-/i- doping of $10^{18}$ and $5\times10^{15}$ cm$^{-3}$ at an excess bias of $V_e=$1.5 V. The excess bias is defined as $V_e=V_a-V_B$, where $V_a$ and $V_B$ are applied reverse bias and breakdown voltage, respectively. For reference, the dark time-dependent current is also plotted in black; the light current curves have colormaps detailed in the upper right schematic, indicating the scanning position of laser input. Fig.~\ref{fig:light_ava_builtup_time}(b) summarizes the avalanche built-up time in dark condition (lines plus symbols) and under light (lines only) as a function of excess bias, extracted from Fig.~\ref{fig:light_ava_builtup_time}(a). Notice that excess bias is calculated based on dark breakdown voltage. For the i-region of 1850 nm under illumination, an earlier breakdown is found at 0.5 V below dark breakdown voltage, thus giving $V_e=-0.5$ V. To fully reveal the spatial variation of light avalanche built-up time, the data in Fig.~\ref{fig:light_ava_builtup_time}(b) can be reorganized to present as a function of laser excitation location, as shown in Fig.~\ref{fig:light_ava_builtup_time}(c). Correspondingly the induced timing jitter can then be obtained by calculating the standard deviation of each curve.

From Fig.~\ref{fig:light_ava_builtup_time}(c), a noticeable trend is that light avalanche built-up time is larger outside the i-region, which can be easily understood since light-generated carriers need to first diffuse to the multiplication region (i-region) before triggering an avalanche. A more interesting observation is that light avalanche built-up time is not always shorter than its dark counterpart, especially when the i-region is thin. Here we highlight a case shown in the lower panel of Fig.~\ref{fig:light_ava_builtup_time}(c) for i-region thickness of 250 nm (i.e., the i-region is located at 1000 to 1250 nm along the NW axis), where the red line indicates the dark avalanche built-up time at $V_e=1.5$ V. It can be seen that for laser excitation beyond 2000 nm, the light avalanche built-up time is longer than the dark one. This is because the minority carriers (i.e. holes) generated at the far end of NW (an n-type region) recombine on their way diffusing to the i-region, given that the maximum hole diffusion length is only about 350 nm (as calculated based on Einstein's relation with carrier mobility, see details in the supplementary material). The excess holes localised in the n-region can also diffuse towards the contact,  contributing to dark current and lowering the light current such that it takes longer for the device to reach avalanche. On the other hand, such observation is not seen for the minority carrier (electron) generated at the p-doped end of NW because of its longer diffusion length ($\sim$ 1 $\mu$m). Overall, the essence is that an excitation spot far away from the multiplication region can directly impair the device response speed.  

\begin{figure}[h]
\includegraphics[width=1\textwidth]{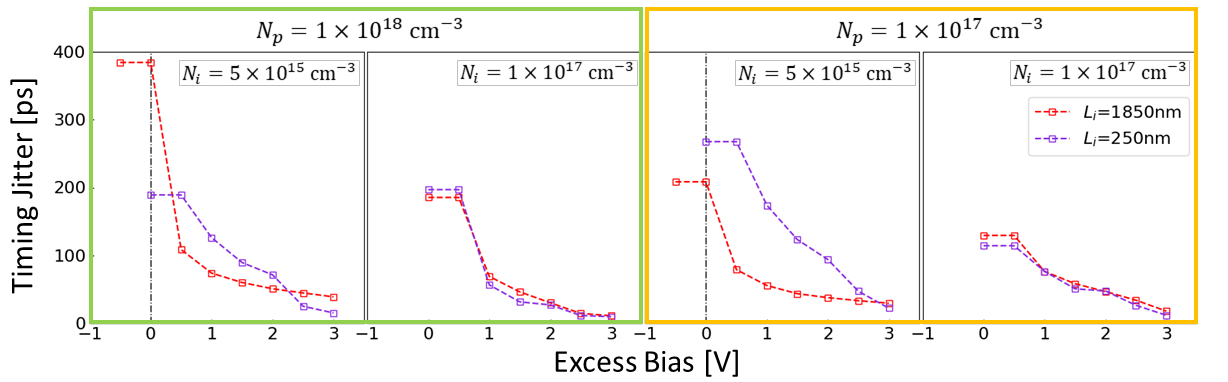}
\caption{ Timing jitter performance for a selection of p-/i- doping and i-region thicknesses. For each p-doping case, we choose the lowest and highest i-doping cases, i.e. $5\times10^{15}$ and $1\times10^{17}$ cm$^{-3}$, with two extreme i-region thicknesses (1850 and 250 nm). 
}  
\label{fig:time_jitter}
\end{figure}

The standard deviation of light avalanche built-up time along the NW, as shown in Fig.~\ref{fig:light_ava_builtup_time}(c), estimates the lower limit of SPAD timing jitter. Given the similar trend observed for different i-region thickness and doping, we summarize the timing jitter performance for a few representative cases in Fig.~\ref{fig:time_jitter}. Timing jitter, similar to avalanche built-up time, decreases as excess bias increases. For low i-doping of $5\times10^{15}$ cm$^{-3}$, a thinner i-region tends to give a larger timing jitter for most bias points. However, as i-doping increases to $1\times10^{17}$ cm$^{-3}$, the difference becomes negligible. Regardless of i-region thickness and doping, for excess bias of more than 2 V, the timing jitter can be less than 50 ps. For high i-region doping at $V_e=3$ V, the timing jitter can be smaller than 10 ps, revealing that single-NW-based SPADs have great potential for high-speed and low-latency applications. 

\begin{figure}[h]
\includegraphics[width=1\textwidth]{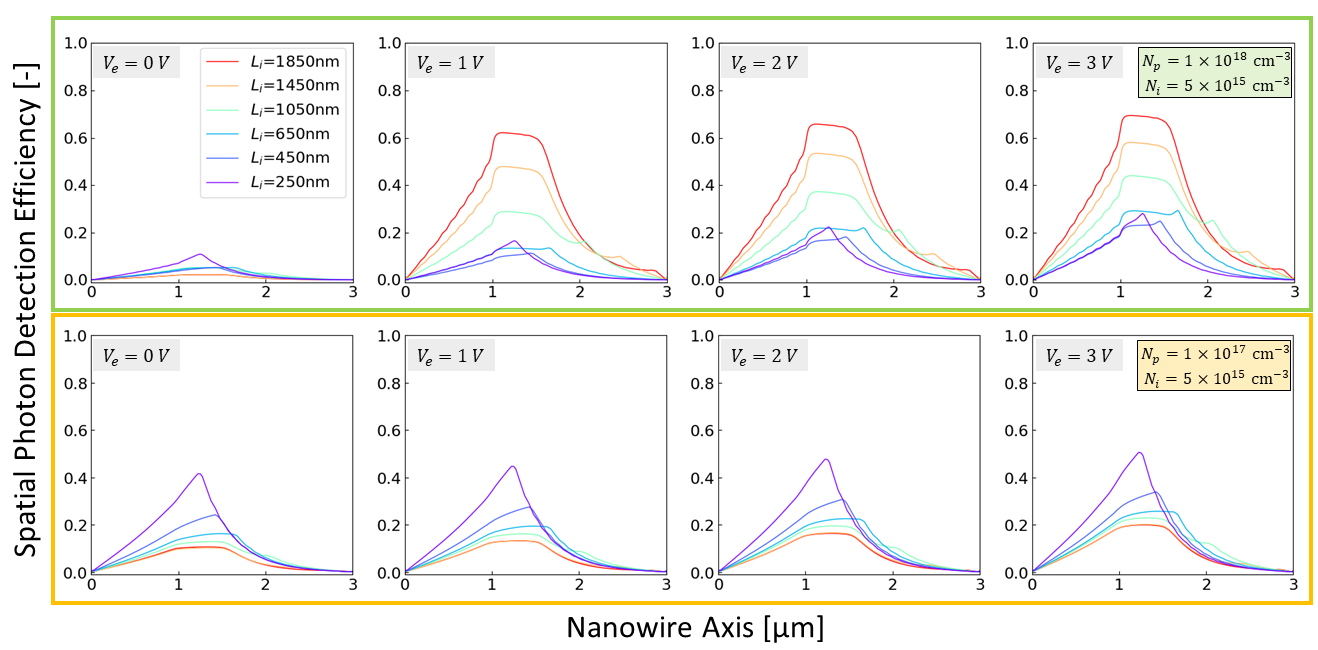}
\caption{Top row: spatial photon detection efficiency PDE($x$) for p-/i-doping of $10^{18}$ and $5\times10^{15}$ cm$^{-3}$ at a range of excess biases. Bottom row: spatial PDE($x$) for p-/i-doping of $10^{17}$ and $5\times10^{15}$ cm$^{-3}$.
}  
\label{fig:spatial_PDE}
\end{figure}

The breakdown voltage, avalanche built-up time and timing jitter discussed above are all extracted directly from the time-dependent DD simulations. For the rest of the SPAD performance metrics, we then used post-processing routines to calculate PDE and DCR. The calculation of detection efficiency also needs a separate quantum efficiency simulation from the fast steady-state DD solver. The spatial PDE($x$) as a function of the NW axis can be calculated as per Eq.~\ref{eq:spatial_PDE}. Fig.~\ref{fig:spatial_PDE} shows two examples of PDE($x$) for high and low p-doping cases with the same i-region doping of $N_i=5\times10^{15}$ cm$^{-3}$ at a range of excess biases. It is easily seen that PDE($x$) increases with a larger excess voltage. More specifically, for the high p-doping case (upper panel), a thicker i-region leads to overall better PDE($x$) along the NW. In contrast, for the low p-doping case, a thinner i-region is preferred. Moreover, for high p-doping, a peak PDE($x$) above  60\% is achieved at $L_i=1850$ nm with higher $V_e$. For the low p-doping case, excess voltage exhibits a less significant effect on PDE($x$); for all i-region thicknesses $L_i$, there is only a slightly increased PDE with increased $V_e$. The highest peak PDE($x$) of $L_i=250$ nm remains above 40\% for all $V_e$.

\begin{figure}[h]
\includegraphics[width=1\textwidth]{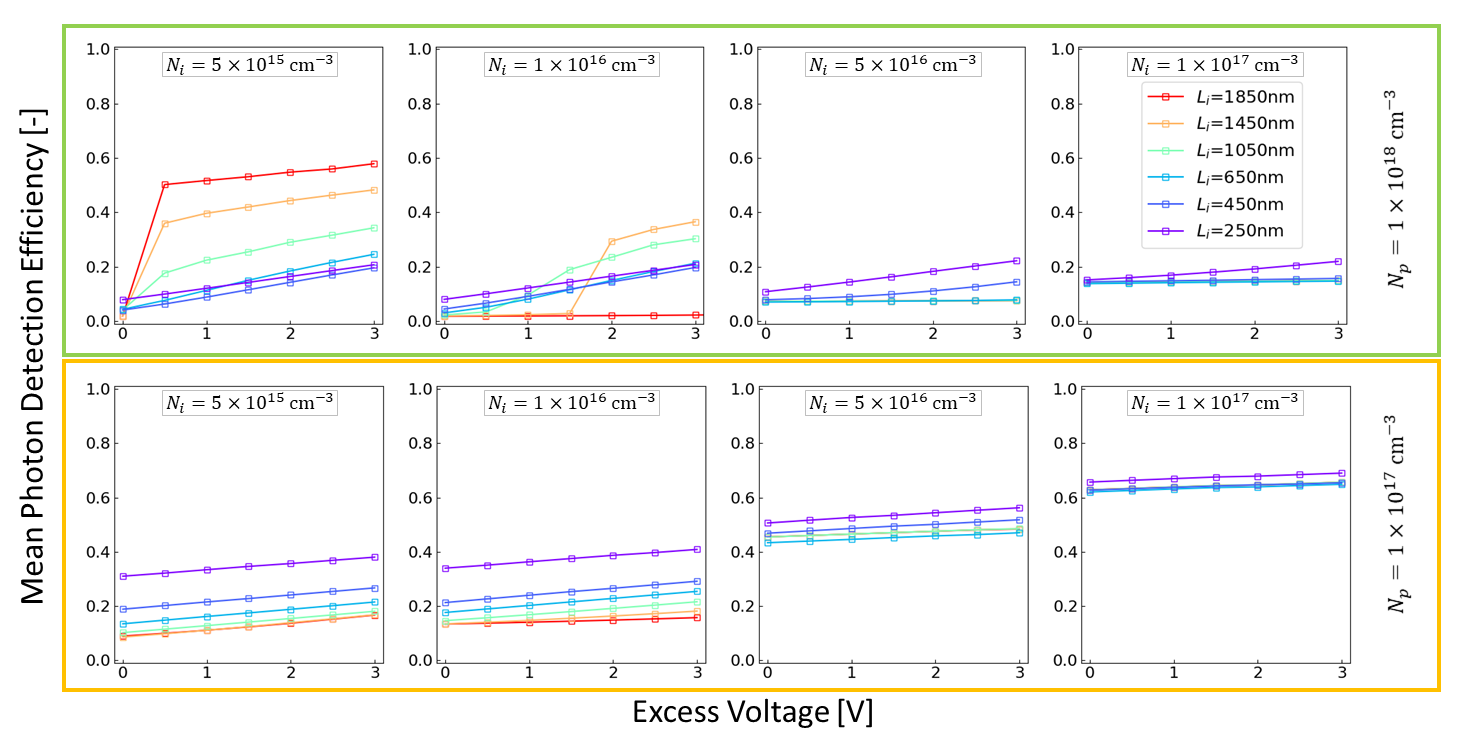}
\caption{Mean photon detection efficiency $\langle \mbox{PDE} \rangle$ is calculated by taking full-width at half-maximum length-averaged PDE($x$) and is plotted against the excess voltage for all doping cases. 
}  
\label{fig:mean_PDE}
\end{figure}

The spatial photon detection efficiency can also be length-averaged within a width of half maximum by Eq.~\ref{eq:length_avg_PDE}, which gives a single-valued mean detection efficiency $\langle \mbox{PDE} \rangle$. Fig.~\ref{fig:mean_PDE} shows that mean detection efficiency $\langle \mbox{PDE} \rangle$ increases with a larger excess bias $V_e$. Also, as i-region doping increases, the difference in $\langle \mbox{PDE} \rangle$ given by different $L_i$ becomes smaller. This is because a larger i-region doping shifts the high-field space-charge region closer to the p-i junction, and for cases such as $N_i=1\times10^{17}$ cm$^{-3}$, the carrier multiplication region is primarily at the p-i junction, shorter than the i-region thickness. Hence, varying $L_i$ has a limited effect on changing the peak field strength, and PDE only increases slightly. Furthermore, suppose p-region doping is also low (such that p-/i-region doping contrast is small). In that case, the high-field multiplication region is essentially always near the p-i junction, regardless of the actual i-region thickness. As a result, a thick i-region is not desirable as it creates an inefficient, low-field quasi-neutral region away from the p-i junction, which is not beneficial for carrier collection and also has a low avalanche triggering probability. This is why for all the low p-doping cases (i.e., the lower panel of Fig.~\ref{fig:mean_PDE}), a thin i-region thickness is preferred. On the other hand, when the doping contrast between the i- and p-region is large, such as in the case of $N_i=5\times10^{15}$ cm$^{-3}$ with $N_p=1\times10^{18}$ cm$^{-3}$, the high-field multiplication region spans the entire i-region, leading to good carrier collection efficiency as well as high avalanche triggering probability throughout the i-region. Consequently, a thick i-region is preferred over a thin one for better PDE performance.

\begin{figure}[h]
\includegraphics[width=1\textwidth]{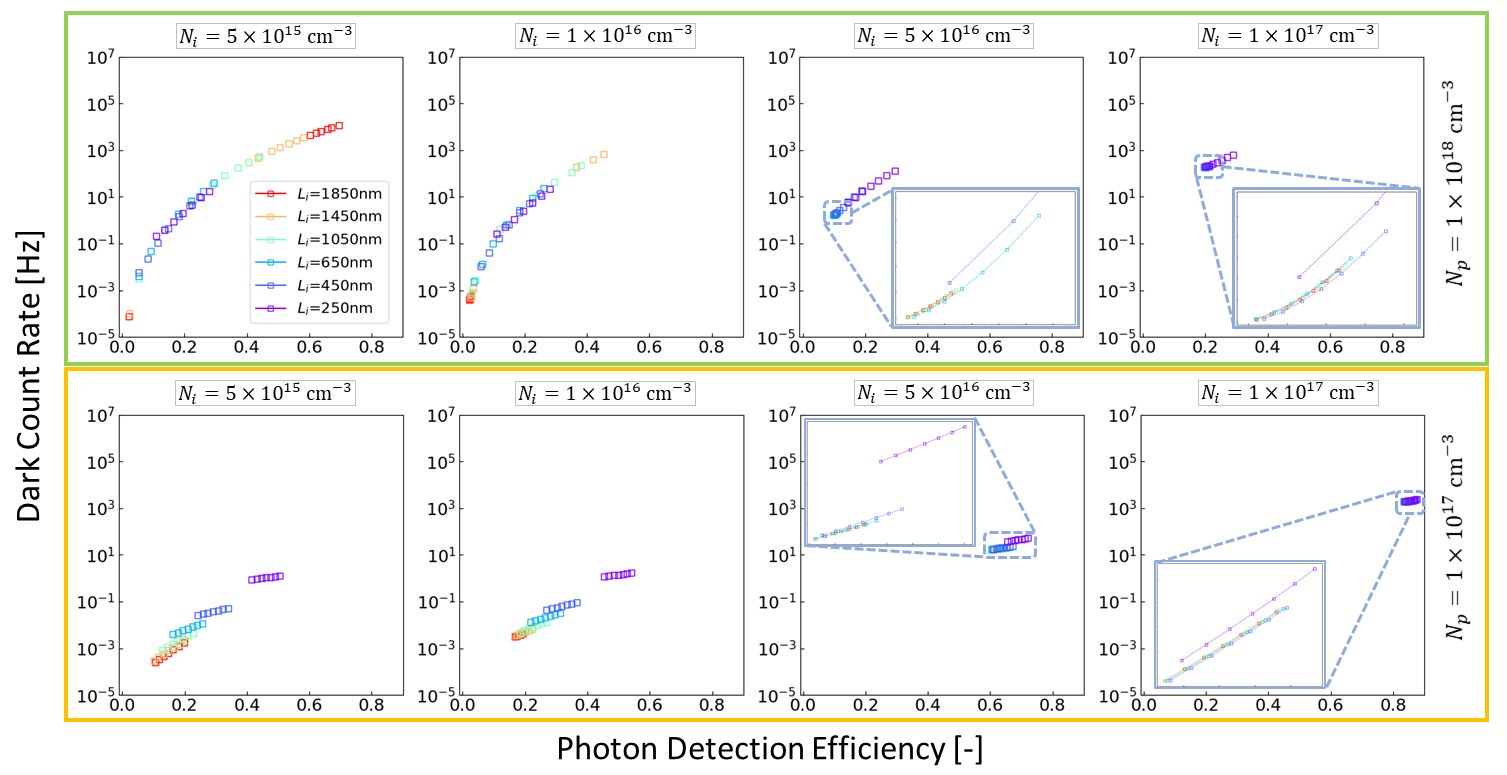}
\caption{Dark count rate versus peak PDE is plotted assuming a high-quality NW with defect density of $N_T=10^{10}$ cm$^{-3}$ for all doping cases. 
}  
\label{fig:dcr_vs_pde}
\end{figure}

Additionally, it should be noted that such length-averaged $\langle \mbox{PDE} \rangle$ only indicates a lower bound of NW-SPAD's peak PDE performance. Nevertheless, NW-SPADs can achieve high mean detection efficiency ranging from 20\% to nearly 60\% for cases of high p-/i-region doping contrast where a thick i-region is preferred (e.g., $N_i=5\times10^{15}$ cm$^{-3}$ with $N_p=10^{18}$ cm$^{-3}$). When the p-/i-region doping contrast is low and a thin i-region is ideal, an even higher efficiency approaching 70\% can be obtained (e.g., $N_i=N_p=10^{17}$ cm$^{-3}$). To compare, the customized cutting-edge planar InGaAs/InP SPDs typically present detection efficiencies of 25\%$\sim$33\%, though at higher operation temperatures of 163$\sim$240 K \cite{Ceccarelli_SPAD_review_AdvQuanTech_2021,HuWeiDa_review_2022}. For array-based NW-SPADs, the light-trapping effect allows for highly concentrated light absorption to be focused precisely around the multiplication region by manipulating array pitch size and NW radius \cite{Wallentin_etal_InP_NWSC_Science_2013}. In this scenario, it is possible for array-based NW-SPADs to achieve the maximum detection efficiency observed in the spatial PDE($x$) profile, where a detection efficiency approaching 90\% can be obtained (see Fig.~\ref{fig:dcr_vs_pde}).

\begin{figure}[h]
\includegraphics[width=1\textwidth]{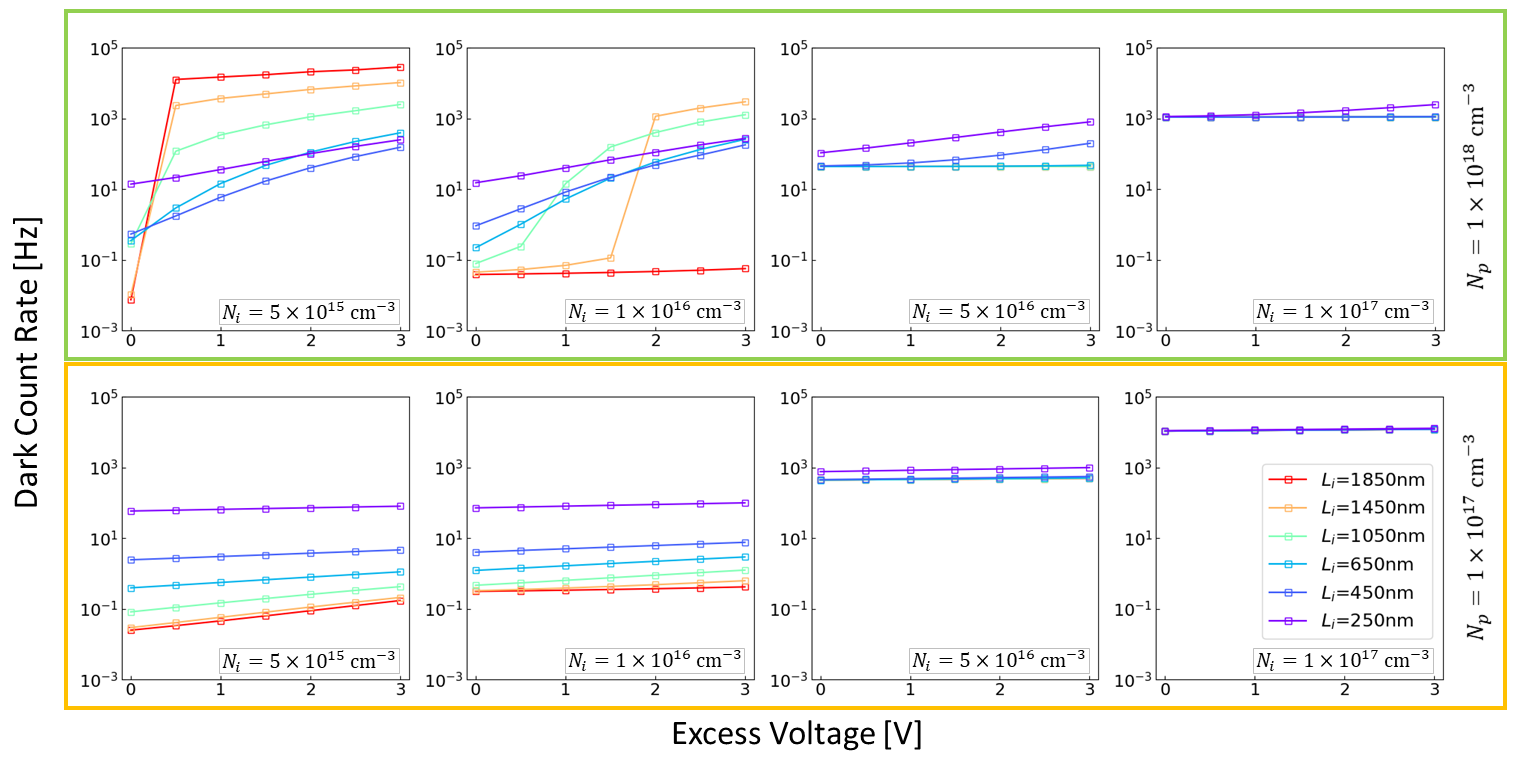}
\caption{The Dark count rate is calculated for a defect density of $N_T=10^{12}$ cm$^{-3}$ for all doping cases.
}  
\label{fig:dcr_vs_Ve}
\end{figure}

To demonstrate the dependence of DCR on i-region thickness and doping, we performed the calculations through the post-processing routines for defect densities of $N_T=$[ $10^{10}$, $10^{11}$, $10^{12}$, $10^{13}$, $ 10^{14}$] cm$^{-3}$. Under the same conditions (i.e., the same $L_i$, $N_i$, $N_p$ and $V_e$), DCR is proportional to the defect density. Fig.~\ref{fig:dcr_vs_Ve} shows an example of computed DCR for all doping cases with a defect density of $10^{12}$ cm$^{-3}$. The computed DCR for the full range of defect densities can be found in Fig.~\ref{fig:dcr_vs_Ve_Appendix} (supplementary material). Based on Eqs.~\ref{eq:G_BBT} and \ref{eq:G_TAT} in the supplementary material, DCR increases with excess bias due to enhanced direct band-to-band and trap-assisted tunneling under a high electric field. However, the rate of increase for higher $V_e$ slows down and eventually becomes much less noticeable as i-region doping increases. This is most evident for the low p-doping case, where the DCR is nearly independent of excess bias. When the doping contrast between the i- and p-doped regions is low, the high-field multiplication region is primarily located at the p-i junction. The over-bias beyond avalanche breakdown does not significantly increase the peak electric field at the junction but merely expands the space-charge region into the neutral region, where the dark carrier generation remains relatively low. This is why the increase of DCR at a larger excess voltage is limited for the cases of low p-doping and high i-doping. On the other hand, in the case of high doping contrast, the high-field space-charge region spans the entire i-doped region. Increasing excess voltage further enhances the peak electric field, where the dark carrier generation is the largest. Hence there is still an apparent increase in DCR as the excess bias increases. For the same reason, it is noted that as i-region doping $N_i$ becomes larger, the DCR dependence on the i-region thickness also becomes much smaller since the high-field multiplication region is always around the p-i junction, regardless of actual i-region thickness.

Overall, by adjusting the i-region doping/thickness and p-region doping, the dark count rate of a single-nanowire SPAD may vary over several orders of magnitude. In particular, for low i-region doping of $5 \times 10^{15}$ or $1 \times 10^{16}$ cm$^{-3}$, a dark count rate less than 1 Hz can be achieved, giving as superior performance as the cutting-edge superconducting nanowire-based SPD \cite{Marsili_SNSPD_NatPhotonics_2013}; such performance, if achieved in the future, is projected to outperform the customized planar InP SPADs, which typically give dark counts ranging from 10 (operated at 163 K) to 26 kHz (at 240 K) \cite{Ceccarelli_SPAD_review_AdvQuanTech_2021,HuWeiDa_review_2022}. Lastly, comparing the trend of dark count rate in Fig.~\ref{fig:dcr_vs_Ve} to that of photon detection efficiency in Fig.~\ref{fig:mean_PDE}, an apparent trade-off between these two metrics can be noted: wherever the detection efficiency is high, a high dark count rate ensues.

To investigate such a trade-off and to highlight the potential of nanowire-based SPADs, Fig.~\ref{fig:dcr_vs_pde} shows dark count rates plotted against the maximum photon detection efficiencies (extracted from spatial PDE($x$) profiles) for all doping cases. We assumed a high-quality NW grown by SA-MOVPE (i.e. a defect density of 10$^{10}$ cm$^{-3}$) that has been demonstrated experimentally  \cite{Ziyuan_IIIV_SingleNW_SolarCell_Review_2018,QianGao_InP_NW_review_2019,JennyWong_NW_review_AM_2020}. Despite a general trend that a higher DCR accompanies a larger PDE,  the degree of such trade-off can be potentially mitigated by properly adjusting the i-region thickness and doping. To guide the material growth and device fabrication for a particular application, for example, aiming at PDE around 70\% (at 700 nm) and DCR between 10 to 100 Hz, one can simply pick an appropriate candidate from eight configurations (of different i-region thickness and i-/p-region doping), shown as eight sub-figures in Fig.~\ref{fig:dcr_vs_pde}. It is noted that a structure of $L_i=250$ nm with the i-/p-region doping of $N_i=5\times 10^{16}$ cm$^{-3}$ and $N_p=1\times 10^{17}$ cm$^{-3}$, respectively, can achieve the design objective. To put this in context, if such a design is realized experimentally, it will closely follow some of the champion and commercial single-photon detectors based on superconducting NWs, whose peak PDE is around 85\% to 95\% with a DCR between $10^2$ to $10^4$ Hz \cite{Korzh_SNSPD_2020,single_quantum_productsheet}, but at a much higher, non-cryogenic temperature (150 K in this work) rather than $\leq$5 K for superconducting NW based detectors. It should be noted that experimentally, it is still challenging to have as delicate control of NW doping as in numerical simulations. Fortunately, based on the simulations, having an i-region doping less or greater than $5\times 10^{16}$ cm$^{-3}$ for $L_i=250$ nm will not substantially compromise the performance. For example, if the i-region ends up being less n-doped, although the peak PDE declines to around 50\%, DCR decreases to less than 10 Hz. On the other hand, a more heavily doped i-region will have a PDE $>80$\% at the expense of increased DCR to 10$^3$ Hz. 

Lastly, it is worth mentioning that the impact of nanowire surface defect states on the NW-SPAD performance was not included due to the 1D device structure studied in this work. The surface states acting as recombination centers in the quasi-neutral regions can capture light-generated carriers as they diffuse towards the multiplication region to trigger an avalanche event, decreasing the device's carrier collection efficiency and eventually leading to a lower photon detection efficiency. Furthermore, carriers captured by the surface defects during the avalanche may be detrapped and induce new avalanche events, i.e., afterpulsing \cite{Farrell_Huffaker_InGaAs_GaAs_SPAD_nanoLett_2019}. Nevertheless, InP nanowires grown by metal-organic vapor-phase or vapor-liquid-solid epitaxy are generally reported with low surface defect densities and surface recombination velocity \cite{HannahJoyce_etal_LowSRV_InPNW_NanoLetts2012,LaPierre_NWSC_review_2013,Ziyuan_IIIV_SingleNW_SolarCell_Review_2018,QianGao_InP_NW_review_2019}. InP nanowire solar cells have been reported with record efficiencies for both epitaxially grown and top-down etched arrays without intentional surface passivation \cite{Wallentin_etal_InP_NWSC_Science_2013,InP_NWSC_TopDown_ACSNano2016}, indicating that surface defect issue may be limited in compromising photon detection efficiency. The effect of surface defects on afterpulsing probability will be investigated in future work. Overall, our simulations indicate that with proper design, it is promising to achieve high-performance NW-based SPADs that can outperform current state-of-the-art detectors based on other technologies.

\section{CONCLUSION}
We have demonstrated an efficient workflow to model a suite of key performance metrics for single nanowire-based single-photon avalanche detectors, which, to our knowledge, has yet to be reported. The workflow incorporates the relatively slow drift-diffusion solver and several fast post-processing routines to maximize computational efficiency. With this, the avalanche breakdown voltage, dark/light avalanche built-up time, deterministic part of timing jitter, dark count rate, and photon detection efficiency can be modeled in an integrated and computationally efficient way. The proposed workflow can effectively serve as a mapping tool between application-driven metrics and a span of design parameters to accelerate the design-fabrication cycle for a high throughput of high-performance devices. Furthermore, our showcase of implementing the workflow onto a single InP nanowire SPAD operating at non-cryogenic temperature reveals the great potential in mitigating the common trade-offs between high detection efficiency/low timing jitter and high dark counts seen in the planar counterparts, thanks to the extremely small active volume of nanowires (i.e., low DCR) and possible tuning of various optical resonant modes for highly localized excitation (i.e., high PDE and low timing jitter). With a proper structural design and material doping, the nanowire SPAD is projected to simultaneously achieve a high detection efficiency of 70\% with a dark count rate less than 20 Hz operated at 150 K. It greatly enhances the freedom in designing avalanche photodetectors and highlights that nanowires are among the most promising candidates for high-performance single-photon detection.

\section*{SUPPLEMENTARY MATERIAL}
Spatial electric field near avalanche breakdown, temperature-dependent physical models, formulations of dark carrier generation rates, and dark count rate of different defect density are given in the supplementary material.

\begin{acknowledgments}
The authors acknowledge the financial support from the Australian Research Council. This research was undertaken with the assistance of resources and services from the National Computational Infrastructure (NCI), which is supported by the Australian Government.
\end{acknowledgments}

\section*{Data Availability Statement}

The data that support the findings of this study are available from the corresponding authors upon reasonable request.

\clearpage
\bibliography{spad_workflow}

\end{document}


\preprint{Supplementary Materials}

\title{Supplementary Materials for:\\ An efficient modeling workflow for high-performance nanowire single-photon avalanche detector}
\author{Zhe Li}
 \email{zhe.li@anu.edu.au;  lan.fu@anu.edu.au}
\author{H. Hoe Tan}
\author{Chennupati Jagadish}
\author{Lan Fu}
 \affiliation{Australian Research Council Centre of Excellence for Transformative Meta-Optical Systems, Department of Electronic Materials Engineering, Research School of Physics, The Australian National University, Canberra ACT 2601, Australia}

\date{August 2023}

\maketitle


\section{Spatial electric field near avalanche breakdown}\label{sec:appendix_efield}
\begin{figure}[h]
\includegraphics[width=1.0\textwidth]{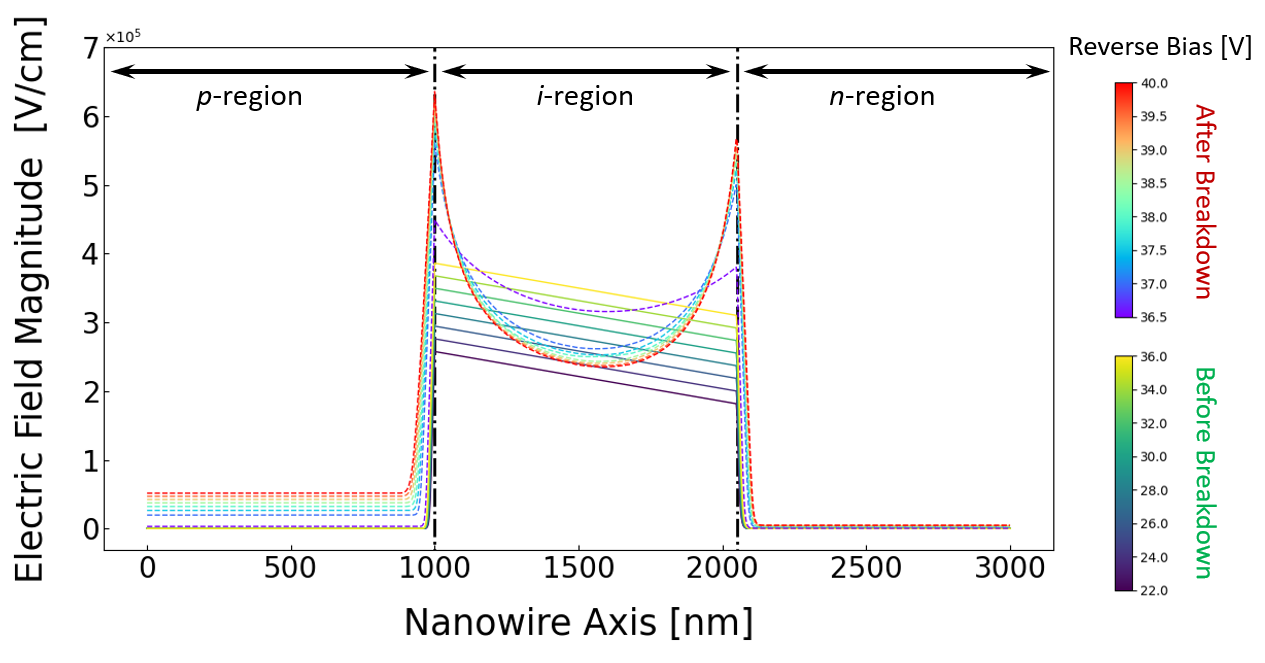}
\caption{Spatial electric field distributions at a range of reverse bias below and beyond avalanche breakdown. The nanowire single-photon avalanche detector (NW-SPAD) has an i-region thcikness of 950 nm with p-/n- and i-segment doping of $1\times10^{18}$ and $5\times10^{15}$ cm$^{-3}$, respectively. The breakdown voltage is -36.5 V. 
}  
\label{fig:efield}
\end{figure}

Fig.~\ref{fig:efield} shows examples of spatial electric field for an nanowire single-photon avalanche detector (NW-SPAD) with i-region thickness of 950 nm and p-/n- and i-segment doping of $1\times10^{18}$ and $5\times10^{15}$ cm$^{-3}$, respectively. Before avalanche breakdown, the electric field in the i-region increases in proportion to reverse bias; once breakdown occurs, the space charge effect becomes evident, where a large number of holes and electrons are accumulated near p-i and n-i junctions, respectively, creating an electric field in the opposite direction to external bias. The resultant electric field screens the field due to external bias, thus leading to a reduction of combined field strength in the central part of i-region. 

\section{Physical models}\label{sec:appendix_phys_model}
For temperature dependent bandgap, a commonly used model is applied in this work \cite{Varshni1967_temperature_bandgap}:
\begin{equation}
	E_g(T_L) = E_{g,0} - \frac{\alpha T_L^2}{\beta+T_L},
\end{equation}
where $T_L$ is the lattice temperature, and the material specific parameters $\alpha$ and $\beta$ are $3.63\times10^{-4}$ eV/K and 162 K, respectively, for InP. The parameter of the band gap at 0 K is $E_{g,0}=1.421$ eV for InP. 

To model temperature and field dependent mobility, we add the temperature-dependent carrier saturation velocity into the commonly used Caughey-Thomas field-dependent model \cite{Canali_MobilityModel_input_parameter_1975}:
\begin{subequations}
\begin{align}
    \mu_n &= \frac{\mu_{n0}}{\left[1+\left(\frac{\mu_{n0}F}{v_{n,\scriptsize{\mbox{sat}}}}\right)^{\gamma_n}\right]^{1/\gamma_n}}, \\
    \mu_p &= \frac{\mu_{p0}}{\left[1+\left(\frac{\mu_{p0}F}{v_{p,\scriptsize{\mbox{sat}}}}\right)^{\gamma_p}\right]^{1/\gamma_p}},
\end{align}
\end{subequations}
where $F$ is the magnitude of a local electric field, $\mu_{n0}$ and $\mu_{p0}$ are the low-field mobility for electron and hole, respectively. We adopt $\mu_{n0}=700 \, \mbox{cm}^2 (\mbox{Vs})^{-1}$ and $\mu_{p0}=100 \, \mbox{cm}^2 (\mbox{Vs})^{-1}$ for InP NWs \cite{Joyce_2013,Vidur_InP_SolarCells_2019}. $\gamma_n$ and $\gamma_p$ are the Canali modification factors that incorporate temperature dependence. However, since the exact temperature dependence is unknown for InP nanowires, we chose not to include the temperature factor and simply adopt $\gamma_n=2.0$ and $\gamma_p=1.0$ as default values based on Refs.~\citenum{Canali_MobilityModel_input_parameter_1975,Ottaviani_correction_to_Canali_mobility_parameter}. 

The temperature dependent saturation velocity for InP is given by Ref.~\citenum{Quay_Vsat_temperature_dependent_2000}:
\begin{equation}
	v_{n,\scriptsize{\mbox{sat}}} = \frac{v_{n,\scriptsize{\mbox{sat}}}(T_L=300\,\mbox{K})}{(1-A_n)+A_n\times(T_L/300\,\mbox{K})}
\end{equation}
for electron, where $A_n=0.68$, and a similar form holds for hole with $A_p=0.3$ \cite{Quay_Vsat_temperature_dependent_2000}. The saturation velocity at $T_L=300$ K also comes from simple Monte Carlo simulation bench-marked with experiments: $v_{n,\scriptsize{\mbox{sat}}}=6.8\times10^{6}$ cm/s and $v_{p,\scriptsize{\mbox{sat}}}=7\times10^{6}$ cm/s at 300 K \cite{Petticrew_2020_modeling_II_InP}.


\section{Dark carrier generation rates}\label{sec:appendix_dcr_generation_terms}

The thermal carrier generation based on Shockley-Read-Hall (SRH) model is given by:
\begin{equation}
	G_{\scriptsize{\mbox{SRH}}} = \frac{n_i}{\tau_n \exp \left[\frac{-(E_T-E_i)}{k_B T}\right]+\tau_p \exp \left[\frac{(E_T-E_i)}{k_B T}\right]},
\end{equation}
where $n_i$ is the intrinsic carrier concentration, $E_T$ is the defect energy level, $E_i$ is the intrinsic Fermi level, $\tau_n$ and $\tau_p$ are the carrier lifetime for electrons and holes, respectively, and $k_B$ is the Boltzmann constant. The temperature dependence comes from $n_i(T)$, $\tau_n(T)$ and $\tau_p(T)$, assuming that defect energy level $E_T$ with respect to $E_i$ does not change with temperature. Note that $E_i$ is also a function of temperature as it depends on band gap $E_g(T)$, and effective densities of states $N_c(T)$ and $N_v(T)$ for conduction and valence bands, respectively. Generally, carrier lifetime increases as temperature decreases; however, its dependence is much less than that of $n_i(T)$, where $n_i(T)$ decreases by four orders of magnitude as temperature decreases from 290 to 200 K. Overall, $G_{\scriptsize{\mbox{SRH}}}$ significantly decreases as temperature decreases because temperature dependence of $n_i(T)$ plays a dominant role. In this work, we assume a defect energy level of $E_T=0.75 E_g$, i.e., $E_T-E_i \approx 0.25 E_g$, to be consistent with the modeling of trap-assisted tunnelling covered in the following part. 

The carrier generation rate due to direct band-to-band tunnelling ($G_{\scriptsize{\mbox{BBT}}}$) used in this work is given by \cite{Donnelly_SPAD_design_consideration,XudongJiang_etal_IEEE_2007}:
\begin{equation}\label{eq:G_BBT}
	G_{\scriptsize{\mbox{BBT}}} = \sqrt{\frac{2 m_r}{E_g}} \frac{q^2 F^2}{4 \pi^3 \hbar^2} \exp\left( - \frac{\pi \sqrt{m_r E_g^3}}{2\sqrt{2}q\hbar F}\right),
\end{equation}
where $F$ is the magnitude of local electric field, and $m_r$ is the reduced mass of the electron effective mass $m_e$ and the light hole effective mass $m_{lh}$ given by $m_r=2(m_e m_{lh})/(m_e+m_{lh})$. In this work, we use $m_e=0.08m_0$ and $m_{lh}=0.089m_0$ \cite{InP_effective_mass_NSM}.

The carrier generation rate due to trap-assisted tunnelling ($G_{\scriptsize{\mbox{TAT}}}$) used in this work is given by: 
\begin{equation}\label{eq:G_TAT}
G_{\scriptsize{\mbox{TAT}}} = \sqrt{\frac{2 m_r}{E_g}} \frac{q^2 F^2}{4\pi^3 \hbar^2}N_T \exp\left(-\frac{\pi\sqrt{m_{lh}E_{B1}^3}+\pi\sqrt{m_{e}E_{B2}^3}}{2\sqrt{2}q\hbar F}\right) \times\left(n_{1,\scriptsize{\mbox{TAT}}}+p_{1,\scriptsize{\mbox{TAT}}}\right)^{-1},
\end{equation}
where 
\begin{subequations}
\begin{align}
    n_{1,\scriptsize{\mbox{TAT}}} &= N_c(T) \exp \left(- \frac{\pi \sqrt{m_e E_{B2}^3}}{2\sqrt{2}q \hbar F}\right), \\
		p_{1,\scriptsize{\mbox{TAT}}} &= N_v(T) \exp \left(- \frac{\pi \sqrt{m_{lh} E_{B1}^3}}{2\sqrt{2}q \hbar F}\right),
\end{align}
\end{subequations}
where $E_{B1}$ is the barrier height of tunnelling from the valence band to the trap, and $E_{B2}$ is the barrier height of tunnelling from the trap to the conduction band. Since there is no existing data available for the exact defect energy level in InP NWs, we adopted $E_{B1} = 0.75 E_g$ used in Refs.~\citenum{Donnelly_SPAD_design_consideration,XudongJiang_etal_IEEE_2007}, which was obtained through the best fit to DCR data for a range of planar InP-based SPADs. The temperature-dependent effective density of states $N_c(T)$ and $N_v(T)$ for InP are given by \cite{Sze_PhysofSemiDev_3rd_ch2}:
\begin{subequations}
\begin{align}
    N_c(T) &= T^{3/2}\times1.1\times10^{14}\,\mbox{cm}^{-3}, \\
		N_v(T) &= T^{3/2}\times2.2\times10^{15}\,\mbox{cm}^{-3}.
\end{align}
\end{subequations}

\section{Dark count rate of different defect density}\label{sec:appendix_dcr_vs_Ve_extra}
\begin{figure}[h]
\includegraphics[width=1\textwidth]{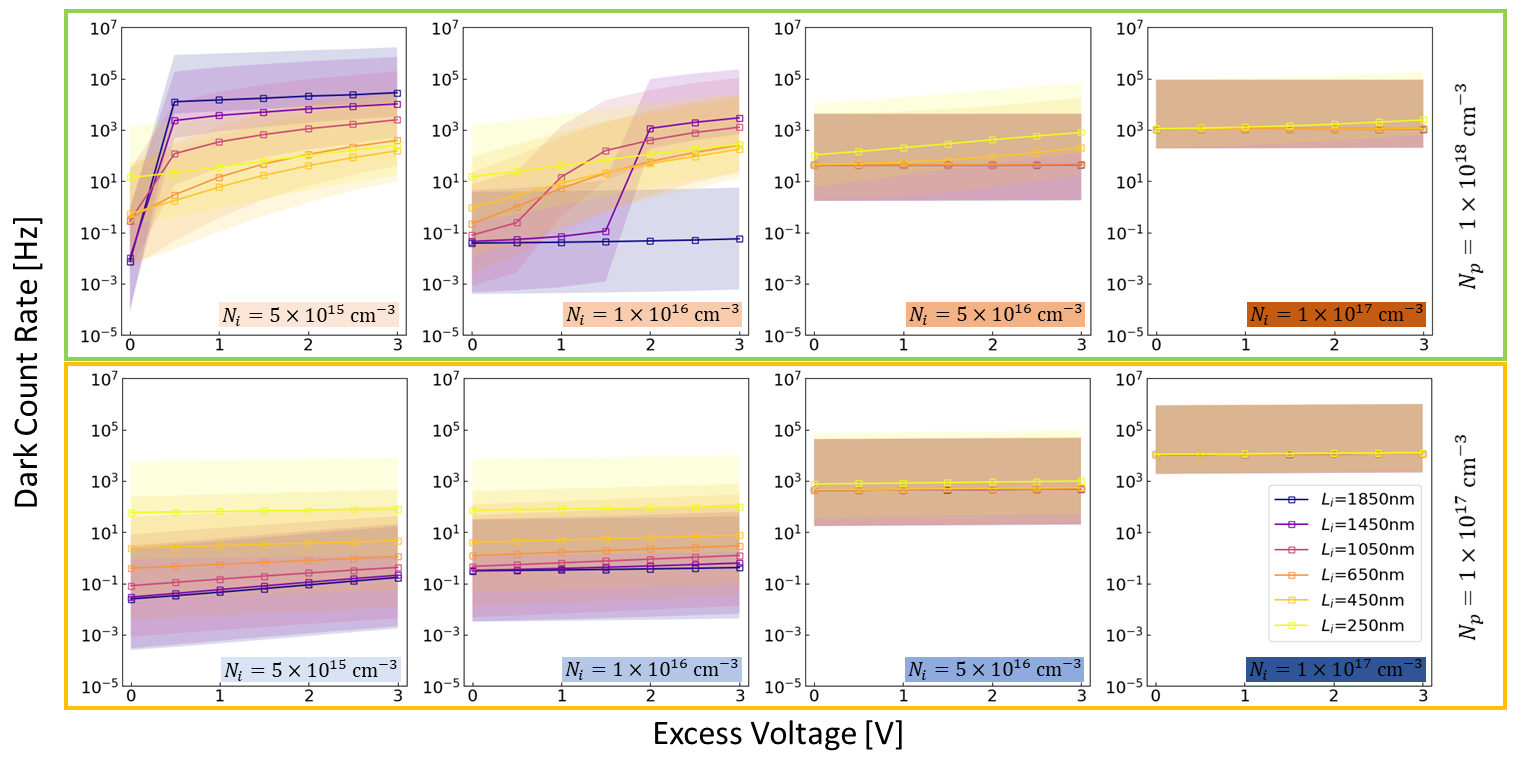}
\caption{The Dark count rate is calculated for a range of defect density $N_T$ and all doping cases, where the solid curve with symbols is for the case $N_T=10^{12}$ cm$^{-3}$ and the associated shaded color depicts DCR ranging from $N_T=10^{10}$ to $10^{14}$ cm$^{-3}$. 
}  
\label{fig:dcr_vs_Ve_Appendix}
\end{figure}

This section presents extra information on the dark count rate calculated for a range of defect density, $N_T=\left[ 10^{10},10^{11},\, 10^{12},\, 10^{13},\, 10^{14} \right]$ cm$^{-3}$, as shown in Fig.~\ref{fig:dcr_vs_Ve_Appendix}.  

\clearpage
\bibliography{spad_workflow}